\documentclass[runningheads]{llncs}
\usepackage{amsmath}
\usepackage{url}
\usepackage{amssymb}
\usepackage[T1]{fontenc}
\usepackage{graphicx}
\usepackage[
    maxbibnames=20,
    backend=bibtex
]{biblatex}
\usepackage{tikz}
\usetikzlibrary{shapes.geometric, arrows}
\tikzstyle{startstop} = [rectangle, rounded corners, minimum width=3cm, minimum height=1cm,text centered, draw=black]
\tikzstyle{process} = [rectangle, minimum width=3cm, minimum height=1cm, text centered, draw=black]
\tikzstyle{decision} = [diamond, minimum width=3cm, minimum height=1cm, text centered, draw=black]
\tikzstyle{arrow} = [thick,->,>=stealth]

\usepackage{orcidlink}
\usepackage{hyperref}
\usepackage{color}

\begin{document}
\title{Weakening Goals in Logical Specifications}
\author{Ben M. Andrew \orcidlink{0009-0009-8910-5899}}
\authorrunning{B.M. Andrew}
%
\institute{University of Manchester, Manchester, UK \\
\email{benjamin.andrew@manchester.ac.uk}}
\maketitle
\begin{abstract}
Logical specifications are widely used to represent software systems and their desired properties. Under system degradation or environmental changes, commonly seen in complex real-world robotic systems, these properties may no longer hold and so traditional verification methods will simply fail to construct a proof. However, weaker versions of these properties do still hold and can be useful for understanding the system's behaviour in uncertain conditions, as well as aiding compositional verification. We present a counterexample-guided technique for iteratively weakening properties, apply it to propositional logic specifications, and discuss planned extensions to state-based representations.

\end{abstract}

\section{Introduction}

Software systems, along with properties that we are interested in proving about them, are often specified in logics such as first-order or temporal logic. Many verification techniques exist for automatically checking whether desired properties hold in a system, but in complex systems that interact with the real world, unexpected environmental conditions or system degradation can cause these properties to no longer hold~\cite{luckcuck2019}.

In these cases traditional formal verification techniques will simply report property violations, leaving us unable to say anything about the system's behaviour. However, for many applications we may be interested in weakened forms of the property that do hold in the system.

As an example, consider a quadrotor drone that we have proven can only safely land when the wind speed is below a certain threshold, and now imagine that this property does \textit{not} hold when one of the rotors has failed. There may still be a weaker version of this property, for example with a lower threshold on the wind speed, that does hold for the degraded drone. Being able to automatically deduce this weakened property may be crucial for regulatory approval or understanding how the system properties change under uncertain conditions.

To be precise, a weakening of a property $P$ is any property that specifies a superset of the behaviours of $P$. For example, we can logically specify the above example as weakening $HighWind \lor LowWind \rightarrow CanLand$ to\linebreak$LowWind \rightarrow CanLand$. (Note that strengthening the antecedent weakens the implication.)

Automatic deduction of weakened properties is also useful in compositional assume-guarantee reasoning~\cite{jones1983}, where our system is an individual component providing guarantees that feed into the assumptions of other components' specifications~\cite{meyer1992}, and aid proofs of the composed system's global properties.

This PhD project currently aims to answer two core research questions:

\begin{description}
\item \textbf{RQ1:} How can a system property, normally holding but invalidated by system degradation or environmental changes, be automatically weakened so that it both holds in the degraded system and is still useful?
\item \textbf{RQ2:} In compositional assume-guarantee reasoning, how does the weakening of a component's guaranteed properties affect other components or system-level properties?
\end{description}

\paragraph{Related Work.} Goal weakening has been explored in requirements engineering~\cite{vanlamsweerde2000}. However, conflicts are only handled between goals because the weakening is done at requirements engineering-time, whereas our approach is concerned with inconsistencies between the requirements and the implementation, which occur at a later stage in the development lifecycle.

Belief revision~\cite{gardenfors1992} is a technique where a belief set is updated when new information conflicts with existing beliefs, removing those that conflict. However, this is a coarse approach that does not weaken the individual beliefs themselves, and so can be overly conservative.

Counterexample-guided techniques have been applied to areas like abstraction refinement~\cite{clarke2000}, inductive synthesis~\cite{alur2013}, and control~\cite{henzinger2003}. However, they have not yet been applied to the problem of weakening goals in specifications.

\section{Counterexample-Guided Weakening}

Our approach finds counterexamples that show that the property doesn't hold in the system, integrates them into the property, and repeats. By integrating counterexamples we iteratively weaken the property until it holds in the system. 

Our algorithm, initially applied to propositional logic, is implemented in OCaml, using the Why3~\cite{filliatre2013} platform with Alt-Ergo~\cite{conchon2018}, a tableau-based solver. The code is hosted publicly on GitHub\footnote{\url{https://github.com/benmandrew/prop-goal-weakening}}.

Not all weakenings of the desired property are useful, as evidenced by the trivial property $\bot$ that any system guarantees. Thus, along with our desired property we also specify a \textit{critical} property $P_C$ that our system must satisfy, as the minimum weakening of $P_D$ that we allow.

Our specifications are triples $\langle A,P_D,P_C \rangle$ of propositional formulae, where $A$ represents the internal structure of the system and the environment, $P_D$ represents the desired property of the system, and $P_C$ represents the critical property of the system. $\langle A,P_D,P_C \rangle$ is well-formed if and only if $P_D$ implies $P_C$, i.e. $P_D \rightarrow P_C$. We begin the proof process by checking whether the iterative algorithm is necessary:

\begin{enumerate}
    \item Check that $A \rightarrow P_D$. If true, then finish successfully with $P_D$ as the property; otherwise,
    \item Check that $A \rightarrow P_C$. If false, then finish unsuccessfully, as our critical property does not hold; otherwise,
    \item Find an intermediate property $P_I$ between $P_D$ and $P_C$ such that $A \rightarrow P_I$.
\end{enumerate}

\noindent
(By $P_I$ being \textit{between} $P_D$ and $P_C$, we mean that $P_D \rightarrow P_I$ and $P_I \rightarrow P_C$, considering propositional formulae to be partially ordered by implication.)

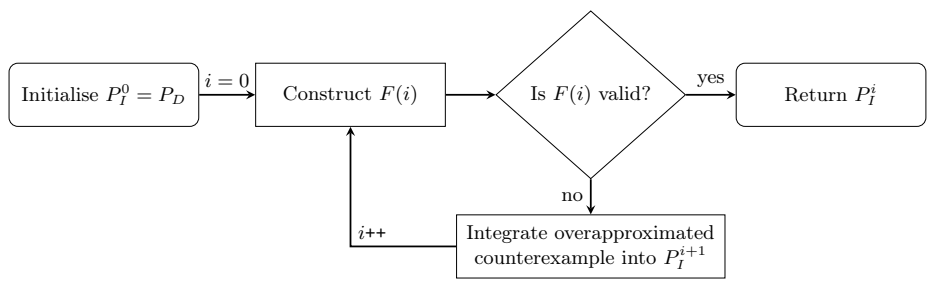
\begin{figure}[ht]
    \centering
    \resizebox{1\linewidth}{!}{%
    \begin{tikzpicture}[node distance=1.9cm]
        \node (start) [startstop] {Initialise $P_I^0 = P_D$};
        \node (construct) [process, right of=start, xshift=2cm] {Construct $F(i)$};
        \node (satsolve) [decision, right of=construct, xshift=1.9cm] {Is $F(i)$ valid?};
        \node (counterexample) [process, below of=satsolve, yshift=-0.5cm] {\begin{tabular}{c} Integrate overapproximated \\ counterexample into $P_I^{i+1}$ \end{tabular}};
        \node (stop) [startstop, right of=satsolve, xshift=1.9cm] {Return $P_I^i$};
        \draw [arrow] (start) -- node[anchor=south] {$i=0$} (construct);
        \draw [arrow] (construct) -- (satsolve);
        \draw [arrow] (satsolve) -- node[anchor=east] {no} (counterexample);
        \draw [arrow] (counterexample.west) -| node[anchor=south west] {$i$\texttt{++}} (construct.south);
        \draw [arrow] (satsolve) -- node[anchor=south] {yes} (stop);
    \end{tikzpicture}}
    \caption{Overview of the algorithm, which iteratively weakens the candidate property $P^i_I$ until it is satisfied by the system representation $A$.}
    \label{algorithm}
\end{figure}

\paragraph{Algorithm.} The algorithm uses a counterexample-guided approach, iteratively computing counterexamples using a SAT solver and integrating them back into the candidate property until it is satisfied. It is detailed below as well as in Fig.~\ref{algorithm}.

The $i$-th candidate property is denoted by $P^i_I$, for $i \in \mathbb{N}$, and we begin by initialising $P_I^0 = P_D$. We then construct a formula $F(i)$ that is a conjunction of the following:

\begin{enumerate}
    \item $A \rightarrow P_I^i$, the candidate property must hold in the system,
    \item $P_D \rightarrow P_I^i$, the candidate property must be weaker than or equivalent to the desired property, and
    \item $P_I^i \rightarrow P_C$, the candidate property must be stronger than or equivalent to the critical property.
\end{enumerate}

\noindent
We check the validity of $F(i)$ using a SAT solver. If $F(i)$ is valid, then $P_I^i$ holds in the system and we are finished. Otherwise, the SAT solver gives us a counterexample model that holds in $A$ but does not in $P_I^i$. This model may contain assignments to \textit{hidden} variables that occur in $A$ but not in $P_D$ or $P_C$. Including them would expose the inner logical workings of the system which may sometimes be desirable, but for the purposes of overapproximating the weakened property, we remove all hidden variables from the counterexample.

From this overapproximated counterexample we construct a formula $C_i$ which is a conjunction of positive or negative propositional variables. For example, the model $\{ X = \text{true}, Y = \text{false}, Z = \text{false} \}$ corresponds to the formula $X \wedge \neg Y \wedge \neg Z$.

The next iteration of the candidate property is then

\centerline{$P_I^{i+1} = P_I^i \lor C_i$}

\noindent
and we repeat the loop by constructing and checking $F(i+1)$.

The algorithm is complete for systems with finite numbers of variables. As each step adds at least one complete interpretation to the candidate property, the number of iterations is bounded by $2^N$, where $N$ is the number of unique propositional variables in $P_D$ and $P_C$ combined.

\paragraph{Example.} We use the example from the introduction concerning a quadrotor drone. Our propositional variables are $R_4$, that all four rotors work; $R_3$, that only three rotors work; $W_H$, that windspeed is high; $W_L$, that windspeed is low; and $L$, that the drone can land.

\noindent
The system is modelled by three assumptions,
\begin{align*}
    A = \underbrace{R_3}_\text{(3a)} \land\; (\underbrace{R_4 \land (W_H \lor W_L) \rightarrow L}_\text{(3b)}) \land (\underbrace{R_3 \land W_L \rightarrow L}_\text{(3c)})
\end{align*}

\noindent
which specify (3a), that only three rotors work (i.e.\ one rotor has failed); and (3b, 3c), the conditions for the drone being able to land. Our desired goal property is $P_D = (W_H \lor W_L) \rightarrow L$, but this is not satisfied by the assumptions, so we must weaken it. (For the purposes of demonstration we let our critical property $P_C = \top$.)

We construct the initial $F(0)$ with $P_I^0 = P_D$, and check for validity with the SAT solver, receiving a negative answer with the counterexample $\neg L \land W_H \land \neg W_L \land R_3 \land \neg R_4$. This counterexample contains the `hidden' variables $R_3$ and $R_4$, and as we would prefer not to expose the inner state of our system, we remove them, resulting in the overapproximated counterexample $\neg L \land W_H \land \neg W_L$. Integrating this into our candidate property results in
\begin{align*}
    P_I^1 &= ((W_H \lor W_L) \rightarrow L) \lor (\neg L \land W_H \land \neg W_L) \\
    &= W_L \rightarrow L
\end{align*}
\noindent
Which is a valid property of the system and so we are done.

\section{Future Work}

To answer \textbf{RQ1}, we are investigating how to extend weakening to properties expressed in state-based specification languages, such as as Deterministic Finite Automata (DFAs), Buchi automata (which commonly correspond with LTL formulae), and Abstract State Machines (ASMs). We are currently exploring how these properties can be automatically weakened, based on the framework of automata learning with the $L^*$ algorithm~\cite{angluin1987}.

Weakening goals is not the only way to weaken a specification: in contract-based reasoning~\cite{meyer1992}, strengthening the corresponding assumption serves the same purpose, and may be a more natural solution for changes in the environment. This will contribute to answering \textbf{RQ2}. It remains to be seen when this would be appropriate, and how exactly it would be done.

It may be more suitable to frame weakening as an interpolation problem~\cite{mcmillan2005} --- that is, finding a suitable interpolant between the desired and critical properties, subject to the constraint of being a valid property of the system. This approach requires investigation. 

\printbibliography

@article{luckcuck2019,
	title = {Formal {Specification} and {Verification} of {Autonomous} {Robotic} {Systems}: {A} {Survey}},
	volume = {52},
	shorttitle = {Formal {Specification} and {Verification} of {Autonomous} {Robotic} {Systems}},
	doi = {10.1145/3342355},
	number = {5},
	journal = {ACM Computing Surveys},
	author = {Luckcuck, M. and Farrell, M. and Dennis, L.A. and Dixon, C. and Fisher, M.},
	year = {2019},
	pages = {100:1--100:41},
}

@incollection{mcmillan2005,
	title = {Applications of {Craig} {Interpolants} in {Model} {Checking}},
	volume = {3440},
	booktitle = {Tools and {Algorithms} for the {Construction} and {Analysis} of {Systems}},
	publisher = {Springer},
	author = {McMillan, K.L.},
	year = {2005},
	doi = {10.1007/978-3-540-31980-1_1},
	pages = {1--12},
}

@article{angluin1987,
	title = {Learning {Regular} {Sets} from {Queries} and {Counterexamples}},
	volume = {75},
	doi = {10.1016/0890-5401(87)90052-6},
	number = {2},
	journal = {Information and Computation},
    publisher = {Elsevier},
	author = {Angluin, D.},
	year = {1987},
	pages = {87--106},
}

@inproceedings{conchon2018,
	title = {Alt-{Ergo} 2.2},
	url = {https://inria.hal.science/hal-01960203},
	author = {Conchon, S. and Coquereau, A. and Iguernlala, M. and Mebsout, A.},
	year = {2018},
    booktitle = {{International} {Workshop} on {Satisfiability} {Modulo} {Theories}},
}

@article{vanlamsweerde2000,
	title = {Handling {Obstacles} in {Goal}-{Oriented} {Requirements} {Engineering}},
	volume = {26},
	doi = {10.1109/32.879820},
	number = {10},
	journal = {IEEE Transactions on Software Engineering},
	author = {{van Lamsweerde}, A. and Letier, E.},
	year = {2000},
	pages = {978--1005},
}

@inproceedings{henzinger2003,
	title = {Counterexample-{Guided} {Control}},
	doi = {10.1007/3-540-45061-0_69},
	booktitle = {Automata, {Languages} and {Programming}},
	publisher = {Springer},
	author = {Henzinger, T.A. and Jhala, R. and Majumdar, R.},
	year = {2003},
	pages = {886--902},
}

@inproceedings{alur2013,
	title = {{Syntax}-{Guided} {Synthesis}},
	doi = {10.1109/FMCAD.2013.6679385},
	booktitle = {{Formal} {Methods} in {Computer}-{Aided} {Design}},
	author = {Alur, R. and Bodik, R. and Juniwal, G. and Martin, M.M.K. and Raghothaman, M. and Seshia, S.A. and Singh, R. and Solar-Lezama, A. and Torlak, E. and Udupa, A.},
    publisher = {IEEE},
	year = {2013},
	pages = {1--8},
}

@inproceedings{clarke2000,
	title = {Counterexample-{Guided} {Abstraction} {Refinement}},
	doi = {10.1007/10722167_15},
	booktitle = {Computer {Aided} {Verification}},
	publisher = {Springer},
	author = {Clarke, E. and Grumberg, O. and Jha, S. and Lu, Y. and Veith, H.},
	year = {2000},
	pages = {154--169},
}

@incollection{filliatre2013,
	title = {Why3 — {Where} {Programs} {Meet} {Provers}},
	volume = {7792},
	booktitle = {Programming {Languages} and {Systems}},
	publisher = {Springer},
	author = {Filliâtre, J. and Paskevich, A.},
	year = {2013},
	doi = {10.1007/978-3-642-37036-6_8},
	note = {LNCS},
	pages = {125--128},
}

@article{jones1983,
	title = {{Tentative} {Steps} {Toward} a {Development} {Method} for {Interfering} {Programs}},
	volume = {5},
	doi = {10.1145/69575.69577},
	number = {4},
	journal = {ACM {Transactions} on {Programming} {Languages} and {Systems}},
	author = {Jones, C.B.},
	year = {1983},
	pages = {596--619},
}

@article{meyer1992,
	title = {Applying `{Design} by {Contract}'},
	volume = {25},
	doi = {10.1109/2.161279},
	number = {10},
	journal = {Computer},
	author = {Meyer, B.},
	year = {1992},
	pages = {40--51},
}

@book{gardenfors1992,
	series = {Cambridge {Tracts} in {Theoretical} {Computer} {Science}},
	title = {Belief {Revision}},
	publisher = {Cambridge University Press},
	author = {Gärdenfors, P.},
	year = {1992},
	doi = {10.1017/CBO9780511526664},
}

\end{document}